\documentclass{jimis-final-en} 

\pdfoutput=1

\usepackage[utf8]{inputenc}
\usepackage{array}
\usepackage{pgfplots}
\usepackage{graphicx}
\usepackage{grffile}
\usepackage[left,modulo,pagewise]{lineno}

\title{ The Problem of Action at a Distance in Networks and the Emergence of Preferential Attachment from Triadic Closure }

\author[*1,2]{Jérôme KUNEGIS}
\author[2,3]{Fariba KARIMI}
\author[2]{SUN Jun}
\affil[1]{University of Namur, Belgium}
\affil[2]{University of Koblenz--Landau, Germany} 
\affil[3]{GESIS -- Leibniz Institute for the Social Sciences, Germany} 
\corrauthor{jerome.kunegis@unamur.be}

\doi{10.18713/JIMIS-140417-2-4}{}
\review{September 6 2016}{April 14 2017}
\publication{2}{2017}

\issue{Graphs \& Social Systems}
\editors{Rosa Figueiredo \& Vincent Labatut}

\begin{document}

\maketitle


\abstract{
In this paper, we characterise the notion of preferential
attachment in networks as action at a distance, and argue that it can
only be an emergent phenomenon -- the actual mechanism by which networks
grow always being the closing of triangles.
After a review of the concepts of triangle closing and preferential
attachment, we present our argument, as well as a simplified model in
which preferential attachment can be derived mathematically from
triangle closing. 
Additionally, we perform experiments on synthetic graphs to demonstrate
the emergence of preferential attachment in graph growth models based
only on triangle closing. 
}


\section{Introduction}

\strut
\vspace{-4ex}

Many natural and man-made phenomena are networks -- i.e., ensembles of
interconnected entities.  To understand such structures is to understand
their creation, their evolution and their decay.  In fact, many models
have been proposed for the evolution of networks, for the simple reason
that a very large number of real-world systems can be modelled as networks. 
Rules for the 
evolution of networks can be broadly classified into two classes: those
postulating local growth, and those postulating global growth.  An
example for a mechanism of local growth is triangle closing: When two
people become friends because they have a common friend, then a new
triangle is formed, consisting of three persons.\footnote{
  In this paper, we use the terms
  \emph{triangle closing} and \emph{triadic closure} exchangably.
  The notion of
  triadic closure has been alluded to multiple times in the history of
  the social sciences; and became mainstream with the work of Mark
  Granovetter~\citeyearpar{b715}.
} This tendency of 
networks to form triangles is a natural model not only for social
networks, but for almost all types of networked data.  For instance, if
Alice likes a movie and Bob is a friend of Alice, Bob might also come
to like that movie.  In this case, the triangle consists of two persons
and one movie.  In general, networks can contain any type of object
being connected by many different types of connections, and thus many
different types of such triangle closings are possible.  We call this
type of growth \emph{local} because it only depends on the immediate
neighbourhood of the two connected nodes; the rest of the network does
not play a role.

\begin{figure}
  \centering
  \subfigure[Triangle closing]{
    \includegraphics[width=0.18\textwidth]{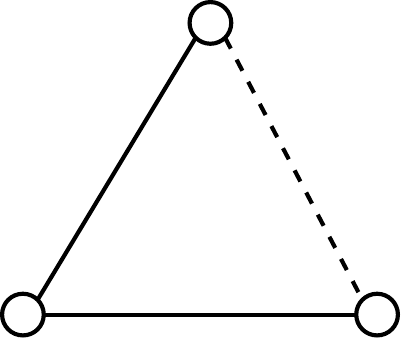}
  }
  \subfigure[Preferential attachment]{
    \includegraphics[width=0.47\textwidth]{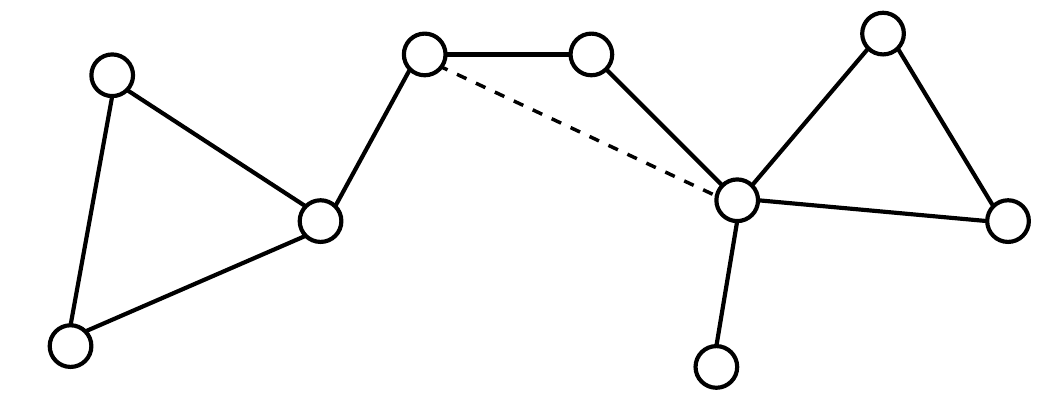}
  }
  \caption{
    \label{fig:illustration}
    The two network growth mechanisms considered in this article:
    triangle closing and preferential attachment.  In both models, new
    edges appear (shown as dashed lines), based on the network
    environment of the current graph. (a)~Triangle closing:  an edge
    is more likely to appear between nodes that have common neighbours,
    (b)~Preferential attachment: A node with higher degree is more likely to 
    receive an edge.
  }
\end{figure}

In contrast to local graph growth rules, there is the phenomenon of
preferential attachment.  When, for  
instance, two people become friends with each other, not because they
have a common friend, or go to the same class, but because one of them or both of them are
popular.\footnote{Preferential attachment, too, is a concept with a long
  history, having been alluded to under multiple names.  See the
  references in \citep{kunegis:preferential-attachment} 
for an account of the early work on it.}  Given a popular
person, i.e.\ with many friends, it is 
more likely that he will be chosen as a friend, than an unpopular
person, all else being equal.  This phenomenon is
referred to as preferential attachment.  Preferential attachment is an
often-used strategy to predict new connections, not only in social
networks: a frequent movie-goer is much more likely to watch a popular
film, than someone who almost never goes out to the movies watching an
obscure film almost nobody knows or has seen.  These types of statements
seem obviously true and indeed they are used widely in application
systems: recommender systems give a big preference to popular movies,
search engines give higher weight to well-connected web pages, and
Facebook or Twitter will make a point to show you pictures that already have many
likes.  In that sense, preferential attachment is true empirically, and
has been verified many times in experiments.  
However, preferential attachment has one problematic property:  It
relies on connecting any two completely unrelated nodes, merely because
of their degree, without considering their interconnections.
Preferential attachment can thus be labelled as ``action at a
distance''.  
For this reason, we argue that
preferential attachment is never a primitive phenomenon, but always a
derived phenomenon, emerging as a result of more basic network evolution
rules, which themselves do not involve action at a distance. 

So, if preferential attachment is not a primitive network evolution
mechanism, which network evolution rules should then be considered as
primitive in our network growth model?  We will
present in this paper arguments for the thesis that
only the principle of triangle closing is fundamental, all forms of
preferential attachment being derived from it. 
To give an
argument in favour of our thesis, we will first review basic notions of
networks and network evolution models, and then review 
preferential attachment, proposing various mechanisms by which
it can arise from triangle closing, a fundamental notion in the
evolution of networks. 
Finally, we perform experiments on synthetic graphs to test to what
extent preferential attachment may emerge from graph growth models that
include only triangle closing and/or random addition of edges. 

\section{Related Work}
The debate over the nature of preferential attachment mechanisms dates
back to the 1960s, when the economist Herbert Simon defended the role of
randomness and the mathematician Benoît Mandelbrot defended the role of
optimisation \citep{barabasi2012network}. The concept of preferential
attachment is also used to explain the nature of scale-free degree
distributions in biological networks such as metabolic networks
\citep{jeong2000large} and protein networks
\citep{jeong2001lethality}. There are various suggestions to explain the
nature of preferential attachment for instance by introducing  hidden variable
models in which nodes possess an intrinsic fitness to other nodes in
unipartite \citep{boguna2003class} or bipartite networks
\citep{kitsak2011}. In a recent \emph{Nature} paper, Papadopoulos and
colleagues proposed a model based on geometric optimisation of homophily
space  
\citeyearpar{papadopoulos2012popularity}. However, in these models, 
triadic closure is not defined as the main principle for the formation
of edges.  

Triadic closure, a tendency to connect to the friend of a friend
\citep{rapoport1953spread}, has been observed undeniably in many social
networks such as friendship at a university
\citep{kossinets2006empirical}, scientific collaborations
\citep{newman2001clustering} and in the World Wide Web
\citep{adamic1999small}. The concept of triadic closure was first
suggested by German sociologist Georg Simmel and colleagues
\citeyearpar{simmel1950sociology} and later on popularised by Fritz Heider
and Mark Granovetter as the theory of cognitive balance in which if two
individuals feel the same way about an object or a person, they seek
closure by closing the triad between themselves
\citep{heider2013psychology}.  
Since the classic preferential attachment model fails to explain the
number of clusters in many social networks, many attempts have been made
to include triadic closure to the model
\citep{holme2002growing,vazquez2003growing}, in 
which nodes connect with certain probabilities based on the principle of
triadic closure.  These works have shown that the scaling law for the
degree distribution and clustering coefficient can be reproduced based
on these models \citep{klimek2013triadic}.
Similarly, models based on random walks as local processes have been
proposed, too, of which triangle closing is a special case
\citep{evans2005}. 

Hence, the scale-free nature of networks and the abundance of
triangles in socio-technical networks beg for
a more fundamental explanation.  Moreover, the observable part of these
systems is not necessarily completely representative for the entire
system.  Networks are generally multi-layered or multiplex, in which some
layers can be hidden or simply not possible to observe
\citep{kivela2014multilayer}.  For instance, the creation of a new Facebook tie
can be caused by attending the same class, sharing the same hobby or
living in a same neighbourhood,  which is hidden from the observable
data.  Consequently, these \emph{focal} points contribute to the tie
creation known as \emph{focal closure} and need to be considered in
modelling realistic networks, as argued by Kossinets and Watts
\citeyearpar{kossinets2006empirical}.  


\section{Networks}
The assertion that networks are to be found everywhere has become a cliché because
it is true.  Social networks, knowledge networks, information networks,
communication networks -- many papers in the field of network science
motivate their use by enumerating fields in which they play a central
role.  Biological networks, molecules, lexical networks, Feynman
diagrams -- hardly a scientific field exists in which networks do not play
a fundamental role.  Instead of giving a hopelessly incomplete
enumeration of examples, we will simply refer the reader to the
introductory section of our Handbook of Network Analysis
\citep{konect:handbook}, in case she wishes to convince herself of
this fact.  In case this is not enough, we may point to the existence of entire
fields of research incorporating the word \emph{network} and synonyms that have emerged
in the last decade: network science \citep{network-science,newman2010networks}, web science
\citep{web-science}, and others \citep{tiropanis2015}.
There are many ways to justify the ubiquitous use of networks as a
model.  As an example, we may consider their use in the field of machine
learning. 
Most classical machine learning algorithms deal with
datasets consisting of data points, each consisting of the same
features.  Mathematically, we may model such a dataset as a set of
points in a space whose dimensions are the individual features \citep{vector-space-model}.
This formalism is very powerful, and still constitutes the backbone of many
machine learning and data mining methods to this day.  The standard
formulation of classification, clustering and other learning problems
all rely on the set-of-points-in-a-space model. However, not all machine
learning problems are well described by the \emph{set of points} model.
While the set of words contained in text documents are well represented
by the \emph{bag of words} model \citep{b359}, a social network is not.  We may try to
represent a social network as a bag of friends, but this
representation is very unsatisfactory:  each person has a set of
friends, but the model does not reflect the fact that a person
contained in one of these bags is the same person as one
\emph{having} a bag of friends.  Thus, the vector space model cannot
find connections such as ``the friend of my friend'' -- it can only find
``a person that has the same friend as me''.  In other words, the vector
space model disconnects the role of \emph{having friends} and that of
\emph{being a friend}.  Instead, the natural way to represent
friendships is as a network.  Using a network model, the symmetry of the
friend relationship is included automatically in the model, and
relationships such as \emph{the friend of my friend} arise as the
natural way to create new edges in the network, i.e., triangle closing.
In fact, we will argue that this is the only way new edges can be created in a network, and
that other models are merely consequences of it, such as preferential
attachment. 

As an additional remark, the terms \emph{network} and \emph{graph} are often used
interchangeably.  Strictly speaking, a network is the real-world object
to be analysed, such as a social network, while a graph is a
mathematical structure used to model it.   

\section{Preferential Attachment}
Preferential attachment, also referred to by the phrase ``the rich get
richer'', or as the Matthew effect, is observed empirically in many
social networks \citep{kunegis:preferential-attachment}.  In fact, the
phenomenon of preferential attachment is known by many other names in
different contexts; see the references within
\citep{kunegis:preferential-attachment} for an account.  In other words,
who has many friends, will get more new friends than who has few.
Movies that have been seen by many people will be seen by more people
than movies that have not.  Websites that have been linked to many times
will receive more new links because of this.  These statements seem
true, and indeed, they are true empirically for many different network
types.

In fact, preferential attachment is the basis for a whole class of
network models.  The most basic of these, the model of
Barabási and Albert (\citeyear{b439}), describes the
growth of a network, which proceeds
as follow:  Start with a small graph, and at each step, add a node, and
connect that node to $k$ existing nodes with a probability proportional
to the number of neighbours for each existing node.  In the limit where
many nodes have been added in that way, the network tends to become
\emph{scale-free}, i.e.\ tends to have a distribution of neighbour counts
that follow a power law.  Since power law degree
distributions are observed in many natural networks, the usual conclusion
is that preferential attachment is correct. 

Preferential attachment is thus undeniably real.  Why then, are we
arguing against it?
The reason is that preferential attachment cannot be a fundamental
driving force for tie creation. 
How are two nodes, completely unconnected from each
other, be supposed to choose to connect with each other?  How can two
completely disconnected nodes even \emph{know} of each other's existence?
This is a fundamental problem with all nonlocal interactions.  For
instance, the classical theory of gravitation as defined and used by
Isaac Newton (\citeyear{newton}) includes nonlocal interactions.  In that
theory, two masses exert a force on each other, regardless of their
position.  While the force decreases with distance, it is always
nonzero, and instantaneous.  The conceptual problem with this type of
interaction had been identified already by Newton himself
\citep{hesse}.  In modern physics, Newton's formalism is replaced by more
precise theories that do not include any action at a distance.  The
theory of general relativity as defined by Albert Einstein  \citeyearpar{einstein} for
instance, only includes local interactions in the form of the Einstein
field equations.  Einstein's general relativity is
thus free from any problematic \emph{action at a distance}, and has been
verified at many experimental scales.  This is also true for
other types of physical interactions -- instead of a force that acts at
a distance between matter particles, quantum field theory models
\emph{bosons} that connect 
particles.  In fact, such interactions can be represented by Feynman
diagrams: graph-like representations of particles in which edges are
particles and nodes are interactions -- any interacting particles must
be connected in one diagram, directly or indirectly.  In this light, we
may interpret preferential attachment as a theory that is true
superficially, but must be explained by an underlying phenomenon.
Specifically, an underlying phenomenon that does not rely on action at a
distance.  As this phenomenon, we propose the known mechanism of
triangle closing.

\section{Triangle Closing}
How do we make new friends?  By meeting the friends of our friends.
This represents a triangle formed by ourselves, our previous friend, and our
new friend.  What if we meet our new friend in another way -- maybe at a
party, or a concert, or at work \ldots\ in any case, there is always
\emph{some} element in common. If we meet our new friend at a party, then
we are both connected to the party, and by modelling the party as a node
in our network, that new friendship is indeed created by the closing of
a person--person--party triangle.  Of course, we may continue to ask how
our connection to the party 
arose.  After all, we did not come to a party randomly.  No, we came to the party because a
friend invited us, or for any other reason, as long as there is some
connection.  This game of connections can be played to any
desired degree of precision.  Maybe we \emph{really} went from door to
door until we found a party with many people.  But then, how did we get
from door to door?  We surely must have started somewhere, likely near to our
home, and have then gone on to the next door, and to the next door, and
so on.  In doing this, we have only followed links:  We are connected to
our home by living there;  our home is connected to the neighbouring
house, which itself is connected to the next house, and so on. 
This example is of course exaggerated, but serves to illustrate the
principle:  in order for a new edge to appear, a path has to exist from
one node to another; this can go over nodes representing any type of
entity, and these nodes may be visible or hidden. 
All in 
all, there is no escaping the principle of triangle closing.  However we
arrived at the party, it must have been by a series of triangle
closings.  

Thus, triangle closing fulfils the expected role as a fundamental mechanism of
network growth, as it is purely local. 
However, we cannot deny the existence of preferential attachment, for which
we must now find suitable explanations. 

\section{Explanations}
In recommender systems, such as those used on web sites that recommend movies to
watch, preferential attachment is often taken as a solution to the cold
start problem.  The cold start problem in recommender systems refers to
the situation in which a user has not yet entered any information about
herself, and thus triangle closing cannot be used to recommend her
anything.  If the user has watched only a single movie, then we can find
similar movies and recommend them.  If a user has added only a single
friend, then we can take movies liked by that friend and recommend them.
But if the user is completely new, as has no friends and no ratings yet,
then this strategy will not work.  How then, do recommender systems give
recommendations to new users?  The solution is simple: they recommend
the most popular items.  If you subscribe to Twitter, you will be
recommended popular accounts to follow.  If you subscribe to Last.fm,
you will be recommended popular music.  For these sites, this strategy
is better than not recommending anything, and in fact is a form of
preferential attachment: Create, or rather recommend, links to nodes
with many neighbours.  How can we interpret this in terms of triangle
closing?  If a node has no connections yet, then surely it cannot
acquire new nodes by triangle closing.  How then will a node ever
acquire new edges, if it starts without neighbours?  The answer is that
a node does not start without any neighbours.  Everything is connected.
A child when it is born does not start without connections; it is
already connected to its parents and to its birthplace.  Likewise, a
user on the Web never starts from scratch: every page has a referrer,
and thus the user can be connected to another website.  Even if the
referring web page is not known, there has to be a referrer.  If a user
types in a URL by hand, she has to have taken it somewhere: maybe a
friend gave it to her, maybe she read it
in a magazine, on a billboard, or on a truck \ldots\ in all cases, the
newly created connection is not created \emph{ex nihilo} -- it is created
by triangle closing.

The explanation for preferential attachment thus lies in hidden nodes:
Nodes that make indirect connections between things, but do not appear
in the model.  On Facebook for instance, many new
friendships are created between people who do not have common friends.
These new friendships seemingly appear without the help of triangle
closing.  However, that is always due to the fact that Facebook does not know
everything.  Some people are simply not on Facebook, which means that if
one meets a new friend through a friend that is not on Facebook and
then connects the new friend via Facebook, then from the point of
view of Facebook a new edge was created without triangle closing.
But
that is only true because Facebook does not know my initial friend.  If
it did, it could correctly infer the new friendship via triangle
closing.  Thus, any two nodes in a network can potentially be linked,
even if they do not share common neighbours \emph{in the network at
  hand}, because they may share a hidden common neighbour.  
The same argumentation applies to hidden nodes that represent
non-actors, such as classes, hometowns, parties, etc. 

In order to justify preferential attachment as an emergent phenomenon,
we must thus derive the mechanism that leads to edges being created
specifically between nodes of high degrees. 
Consider a network, for instance a social
network.  Call this the known network.  Then, consider a certain number of
nodes outside of that 
network, that are connected at random to the nodes in the known
network.  Call these the unknown nodes.   How many common neighbours do
two members of the known network 
have outside of the known network?  Without knowing the distribution of
hidden edges, this question cannot be answered.  But consider that
triangle closing acts not only on known--unknown--known paths, but
also on known--known--unknown paths.  Starting with an equal
probability for all known--unknown edges, performing triangle closing
will lead to the creation of known--known--unknown triangles.  The newly
created known--unknown edges can then be combined with other
unknown--known edges to perform, again, triangle closing, leading to new 
known--known edges.  The result are new edges in the observed social
network, with a probability proportional to the number of the initial
known node's neighbours.  
Thus, preferential attachment emerges as a necessary consequence of
iterated triangle closing, if hidden nodes are admitted.
The next section will make this heuristic argument precise. 

\section{Derivation}
This section gives an exemplary derivation of a simplified model that we
introduce to illustrate that preferential 
attachment arises as a consequence of triangle closing in the presence
of hidden nodes.  The given scenario is very general and may be generalised
easily for instance by considering multiple node types or multiple edge types.  
In this model, we distinguish two types of nodes:  visible nodes in the set $V$,
and hidden nodes in the set $W$.  We will assume that there is a given,
fixed number of visible nodes $|V|$, and a possibly very large number of
hidden nodes $|W|$.  In particular, we will consider the limit $|W|
\rightarrow \infty$. 

Let $G=(V \cup W, E)$ be the graph representing the complete system, in
which $V$ is the set of visible nodes, and $W$ the set of hidden nodes.
Additionally, $E$ is the set of edges connecting nodes in $V$ with nodes
in $W$.  While we assume that the individual edges in $E$ are hidden,
the degree of the nodes in $V$ is not hidden.  In other words, the
number of edges of $E$ incident to each node in $V$ is known.  
Edges between nodes
in $V$ will not be considered. 
Likewise, edges between nodes in $W$ need not be
considered, since they do not contribute to the degree of nodes in $V$.
Thus, the considered network $G$ is bipartite.  
We will use the convention that $n = |W|$, and the degree of a node $u$ is
denoted by $d(u)$. 
We now assume that the graph $G$ will receive new edges according to the
principle of triangle closing. 
Thus, two nodes in $V$ will connect with a probability
proportional to the number of common neighbours they have.  
Seeing only nodes in $V$ and their degree, preferential attachment can then
be observed as described in the following. 

In order to make our derivation, we need to make two assumptions:
\begin{itemize}
  \item The triangle closing process is random in the sense that new
    edges are added between any possible node pairs with equal
    probability. 
  \item The typical degree of nodes is significantly smaller than the
    number of nodes, i.e., $d(x) \ll n$.  This is precise when $n$ goes
    to infinity.  
\end{itemize}

Let $u,v \in V$ be two nodes of the network.  Under the assumption that
the edges are distributed randomly in the graph, the probability $p$
that $u$ and $v$ are connected can be derived combinatorically by
considering the number of configurations in which the two nodes do not
share a common neighbour.  Given that $u$ and $v$ have degree $d(u)$ and
$d(v)$ respectively, the total number of configurations for the edges
connected to the nodes is
\begin{align}
  {{n} \choose {d(u)}} {{n} \choose {d(v)}}.
\end{align}
Out of those, the number of configurations in which the neighbours of
the two nodes are disjoint is given by
\begin{align}
  {{n} \choose {d(u)}} {{n - d(u)} \choose {d(v)}}.
\end{align}
Thus, the probability that the two nodes share a common neighbour is
given by 
\begin{align}
  p &= 1 - \frac
  {{{n} \choose {d(u)}} {{n - d(u)} \choose {d(v)}}}
  {{{n} \choose {d(u)}} {{n} \choose {d(v)}}}
  = 1 - \frac
  {{{n - d(u)} \choose {d(v)}}}
  {{{n} \choose {d(v)}}}.
\end{align}
We now use the falling factorial to express binomial coefficients, i.e.,
\begin{align}
  n^{\underline{a}} = n(n-1)(n-2)\cdots(n-a+1). 
\end{align}
The falling factorial has the property that in the limit where $a$ is
constant and $n$ goes to infinity, we have
\begin{align}
  \lim_{n \rightarrow \infty} \frac {n^{\underline{a}}}{n^a} = 1
\end{align}
and also,
\begin{align}
  {n \choose a} = \frac {n^{\underline{a}}} {a!}, 
\end{align}
and thus
\begin{align}
  p = 1 - \frac
  {(n - d(u))^{\underline{d(v)}} d(v)!}
  {d(v)! n^{\underline{d(v)}}} 
  = 1 - \frac
  {(n - d(u))^{\underline{d(v)}}}
  {n^{\underline{d(v)}}}.
\end{align}
In the limit when $n$ goes to infinity we may thus assume that
\begin{align}
  p = 1 - \frac
  { (n - d(u))^{d(v)} }
  { n         ^{d(v)} }
  = 1 - \left( 1 - \frac {d(u)} {n} \right)^{d(v)}
\end{align}
and using again the limit $n \rightarrow \infty$, and the property that
in the limit where $\varepsilon$ goes to zero, $(1-\varepsilon)^k$ goes to
$(1-k\varepsilon)$, 
\begin{align}
  p = \frac {d(u)d(v)} {n}. 
\end{align}
It thus follows that $p \sim d(u)d(v)$, i.e., the probability of the
nodes $u$ and $v$ being connected is proportional to both $d(u)$ and $d(v)$.
Thus, we find that preferential attachment is a consequence of the triangle closing
model. 
Preferential attachment itself then leads to a scale-free degree
distribution, as per Barabási and Albert \citeyearpar{b439}. 

\section{Experiments}
In this section, we give empirical evidence for the emergence of
preferential attachment in graph growth models that do not include it. 
In the experiments, we generate synthetic networks via a random growth
process that does not include preferential attachment, as well as
using random growth processes that do include preferential attachment.
In all generated networks, effects of preferential attachment are then
measured empirically. 
All generated
networks have 1,000 nodes and 10,000 edges, and are undirected, loopless, and
do not allow multiple edges.  In all cases, the graphs are generated by
starting with a graph of 1,000 nodes and without edges, and adding edges
one by one. 
For each edge that is added, one of the following three methods is
chosen at random: 
\begin{itemize}
  \item \textbf{Random}:  With probability $p_{\mathrm{r}}$, an edge is added randomly between two
    unconnected nodes.  All pairs of distinct unconnected nodes are chosen with
    equal probability. 
  \item \textbf{Triangle closing}:  With probability $p_{\mathrm{tc}}$, among all unclosed triads, one is
    chosen randomly with equal probability, and the third edge is
    added.  An unclosed triad is a triple of nodes $(u,v,w)$ such that
    $(u,v)$ and $(u,w)$ are connected, but $v$ and $w$ are not connected.
    If chosen, the triangle is completed by adding the edge $(v,w)$.  If
    no unclosed triads are present, an edge is added at random as
    described in the previous case.
  \item \textbf{Preferential attachment}: With probability $p_{\mathrm{pa}}$, a node is chosen with a
    probability proportional to the node's current degree.  Then,
    out of all nodes not connected to that node, one is chosen randomly
    and with equal probability, and an edge is added between the two
    selected nodes.  If there is not at least one unconnected pair of nodes with nonzero degree,
    an edge is added at random as described in the first case. 
\end{itemize}
In each experimental trial, the three probabilities are chosen such that 
$p_{\mathrm r} + p_{\mathrm{tc}} + p_{\mathrm{pa}} = 1$.  Each of these probabilities
is varied from 0 to 1 in increments of 1/11, excluding the case
$p_{\mathrm r}=0$ in order to avoid the runaway case of an individual node
accumulating all edges.\footnote{In the degenerate case of $p_{\mathrm r}=0$,
  almost all edges will be attached to a single node in the $n
  \rightarrow \infty$ limit.} 

\begin{figure}
  \centering
  \includegraphics[width=0.6\textwidth]{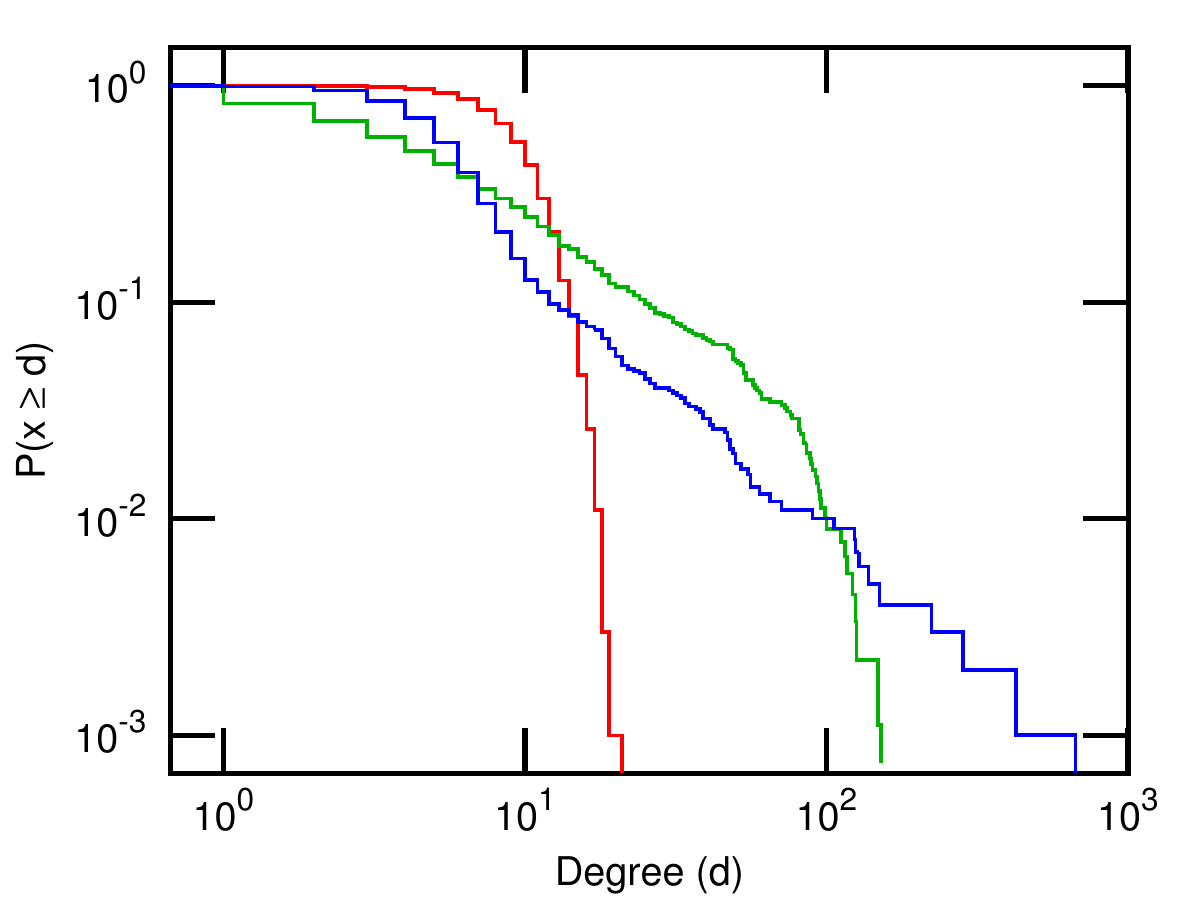}
  \raisebox{3cm}{\includegraphics[width=0.15\textwidth]{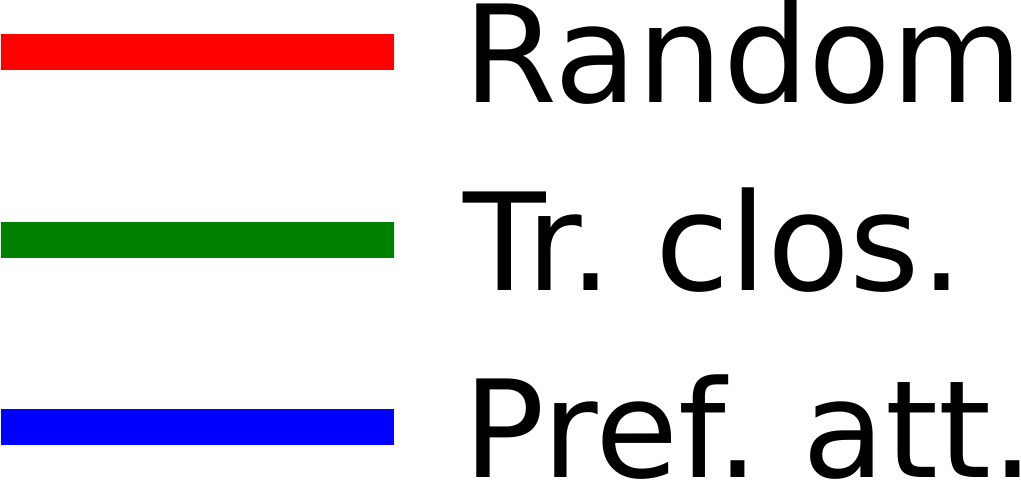}}
  \caption{
    \label{fig:comparison}
    The cumulative degree distribution for the three extremal generated
    networks. 
  }
\end{figure}

First, in order to verify whether a graph created by the process of
triangle closing display scale-free behaviour,  
we compare the generated distribution of the triangle closing case with
the degree distributions for the random and preferential attachment
cases.\footnote{As described in the previous paragraph, the cases of pure
  triangle closing and preferential attachment also include 1/11 of
  edges based on random assignment.}  All three degree distributions are shown in
Figure~\ref{fig:comparison}.   
In the plot, several observations can be made.  The degree distribution
for the triangle closing case displays power law-like behaviour over
multiple orders of magnitude, from the smallest degrees of one, to
approximately one hundred.  
While the networks generated by triangle closing and preferential
attachment have similar power-like degree distribution, both with
comparable exponent, we must note that the maximum degree in the
preferential attachment case is larger than in the triangle closing
case. 
However, the triangle closing model displays a power law degree
distribution with exponential cut-off that has been observed in many
real networks due to finite size effect
\citep{boguna2004cut,clauset2009power}. 
For comparison, the
preferential attachment case also displays power law-like behaviour,
although not for very small degrees (under about 10), and additionally
has a well-defined long tail.  The purely random case leads to a degree
distribution that shows no scale-free behaviour.  

We measure the equality of the distribution of edges, or its opposite,
its skewness, as the primary consequence of the preferential attachment
process.  As a measure, we use the Gini coefficient of the degree
distribution, as defined in \citep{kunegis:power-law}.  The Gini
coefficient is zero when all nodes have equal degree, and attains its
theoretical maximum of one when all nodes except a single one have
degree zero.\footnote{Since an edge always connects two nodes, the
  actual maximum is attained in star graphs, in which all edges 
  attach to a single node, and other nodes have a degree of zero and
  one.  In the large-graph limit, the Gini coefficient in such graphs
  tends to one.}  

\begin{figure}
  \includegraphics[width=0.8\textwidth]{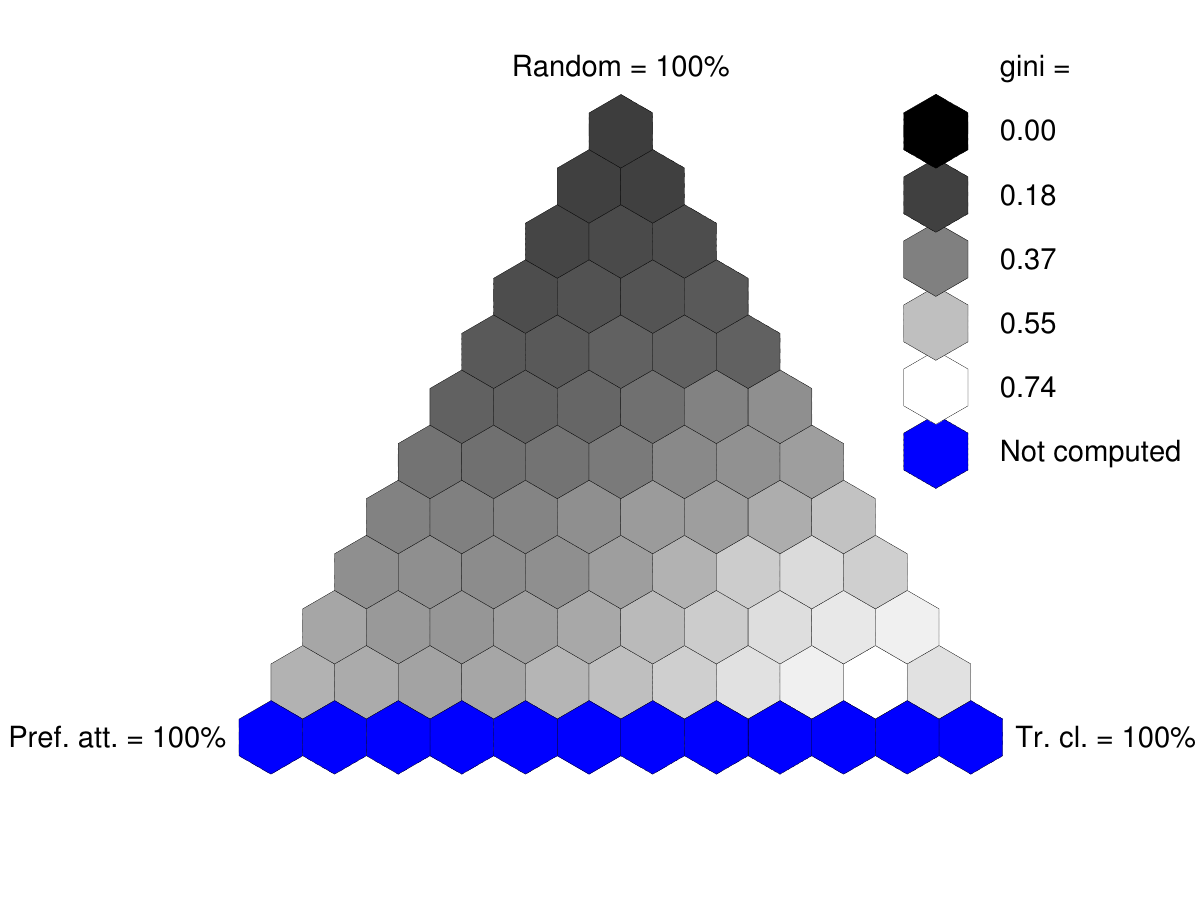}
  \caption{
    \label{fig:simplex-gini}
    Experimental results:  Each cell shows one experimental run with a
    different probability of adding each edge at random (top), via
    triangle closing (bottom right), and preferential attachment (bottom
    left).  The bottom row was not executed due to the tendency of
    models with random edges to attach all edges to a single node,
    giving values of the Gini coefficient very close to the theoretical
    maximum of one. 
  }
\end{figure}

The experimental results are shown in Figure~\ref{fig:simplex-gini}. 
In the triangle shown in the figure, the top-to-bottom-right edge shows the
cases in which preferential attachment is excluded, while the
bottom-left corner (\emph{Pref.\ att.\ = 100\%}) represents the case of exclusive preferential
attachment.  As expected, the 100\% random case results in an
Erdős--Rényi graph in which the degrees have a Poisson distribution, and
thus a very uniform number of edges over all nodes, giving a small Gini
coefficient of 17.7\%.  The pure preferential attachment case gives a
higher value of about 51.5\%.\footnote{In this and all subsequent cases
  labelled as \emph{pure}, the method in question has a probability of
  $p_{\mathrm{tc},\mathrm{pa}}=10/11$ while a random edge is added with a probability of
  $p_{\mathrm r} = 1/11$.}  The pure triangle closing method results
in a value of the Gini coefficient of 65.1\%, a value similar to (and
even superior to) the value in the pure preferential attachment
case.  Thus, it is indeed the case that a skewed degree distribution is
generated by a purely local process of triangle closing, without the need for explicit
preferential attachment.  We note also that preferential attachment is
observed even though the number of nodes in the network ($n=1{,}000$) is
relatively small when compared to the theoretical model described in the
previous section in which the limit $n \rightarrow \infty$ is taken. 

\section{Discussion}
Our experiments have allowed us to observed that triangle
closing leads to skewed and scale-free degree distributions. 
However, the status of a mechanism as \emph{fundamental} is not clear cut.  When
a phenomenon is explained by another, more fundamental phenomenon, 
we can consider it as derived.  But how can we be sure that a phenomenon
is not explained by a more basic phenomenon?  What does it mean for a
phenomenon to be fundamental?  Just as physics cannot declare one theory
to be final, we cannot declare one network growth mechanism
to be final.  Thus, individual instances of triangle closing can for
instance be explained by several layers of triangle closing, just as in
physics a direct interaction can be explained by a new mediating
particle.  In the end however, this applies only to specific instances
of triangle closing, as it replaces them with other, more detailed
instances of triangle closing.  Thus, triangle closing \emph{does} play a
fundamental role in growing network models, only that it cannot always
be derived which three nodes are taking part in it, as one of the three
nodes is often hidden.  In the end, the only judge of
the validity of a model remains the experiment, and in practice, used
models do not have to be fundamental -- recommenders and information
retrieval systems have had enough success by applying 
preferential attachment directly.

As mentioned in the introduction, triangle closing is itself a general
phenomenon that not only applies to pure social networks, but also to
other types of networks.  In the case of property networks, i.e.,
networks containing edges between persons and the properties they have,
triangle closing can be identified with the concept of homophily, i.e.,
the concept that friends tend to be similar.  As an example, the fact
that two smokers become friends can be modelled as the closure of the
(person~A)--(colleague~+~smoker)--(person~B) triangle, in which
``colleague~+~smoker'' is a non-person 
node of the network representing the property of \emph{being a colleague
  and a smoker}.
Thus, the fact that friends of smokers are more likely to be smokers
too (a classical example of homophily) can be analysed as a form of
triangle closing in a graph that is not purely a social network, as it
contains non-person nodes.  Homophily is thus consistent with the view
that triangle closing is fundamental \citep{shalizi2011homophily}. 

The problem posed in this paper can be generalised to other graph growth
mechanisms.  For instance, we may ask whether assortativity (the
tendency of connected nodes to have correlated degrees) or community
structures emerge from triangle closing alone.  In the case of community
structures, triangle closing trivially plays a role, as triangle closing
by construction leads to tightly connected graphs.  As for
assortativity, the fact that both assortativity (a positive correlation
between degrees) and dissortativity (a negative correlation between
degrees) have been observed in social networks points to the fact that a
single model such as triangle closing cannot (and is not expected to)
explain all properties of a social network, and other phenomena must be
at work, which may or may not be local. 

\bibliographystyle{jimis-en}
\bibliography{nopref,ref,kunegis,konect}

\end{document}